# Ethics and Persuasion in Reinforcement Learning from Human Feedback: A Procedural Rhetorical Approach

Shannon Lodoen
Humanities and Communication
Embry-Riddle Aeronautical University
Prescott, United States
lodoens@erau.edu

Alexi Orchard
Technology and Digital Studies Program
University of Notre Dame
Notre Dame, United States
aorchard@nd.edu

*Abstract*—Since 2022, versions of generative AI chatbots such as ChatGPT and Claude have been trained using a specialized technique called Reinforcement Learning from Human Feedback (RLHF) to fine-tune language model output using feedback from human annotators. As a result, the integration of RLHF has greatly enhanced the outputs of these large language models (LLMs) and made the interactions and responses appear more "human-like" than those of previous versions using only supervised learning. The increasing convergence of human and machine-written text has potentially severe ethical, sociotechnical, and pedagogical implications relating to transparency, trust, bias, and interpersonal relations. To highlight these implications, this paper presents a rhetorical analysis of some of the central procedures and processes currently being reshaped by RLHF-enhanced generative AI chatbots: upholding language conventions, information seeking practices, and expectations for social relationships. Rhetorical investigations of generative AI and LLMs have, to this point, focused largely on the persuasiveness of the *content* generated. Using Ian Bogost's concept of procedural rhetoric, this paper shifts the site of rhetorical investigation from content analysis to the underlying *mechanisms* of persuasion built into RLHF-enhanced LLMs. In doing so, this theoretical investigation opens a new direction for further inquiry in AI ethics that considers how procedures rerouted through AI-driven technologies might reinforce hegemonic language use, perpetuate biases, decontextualize learning, and encroach upon human relationships. It will therefore be of interest to educators, researchers, scholars, and the growing number of users of generative AI chatbots.

## I. Introduction

As of 2025, generative AI is used widely through chatbot interfaces such as ChatGPT (OpenAI), Claude (Anthropic), and Google Gemini, among others. Large language models (LLMs), the basis of these chatbot interfaces, are built using a combination of supervised learning, reinforcement learning, and other natural language processing techniques. Since 2022, versions of chatbots such as ChatGPT and Claude have been trained using a specialized reinforcement learning technique called Reinforcement Learning from Human Feedback (RLHF) to fine-tune the model output using feedback from human annotators.

This paper analyzes the inherent persuasive aspects "baked into" LLMs through their RLHF-enhanced training process. We examine how this disembodied, seemingly un-authored text nonetheless reflects human motives, values, and ideals that have been embedded here, as they are in all technological objects, systems, and structures [1], [2]. We also consider the question of user experience and how arguments to users are subtly constructed through interactions with LLMs in different contexts.

To this point in time, research at the intersection of rhetoric and generative AI has been largely limited to analyzing, producing, or testing various prompts or generated outputs. Topics of investigation include "Socratic" chatbots for pedagogical purposes [3], [4], [5], concerns over perceptions of accuracy and truthfulness [6], [7], [8], and testing of various LLMs' capabilities to produce rhetorically effective, i.e. persuasive, content [9], [10], [11], [12], [13]. Thus, this paper opens a new avenue for future, procedurally oriented critiques of artificial intelligence and other emerging technologies.

We begin by providing a literature review and general overview of RLHF and its role in enhancing LLMs. We then turn to rhetoric—an area of inquiry spanning over two millennia concerned with persuasion and communication —to assess the implications of LLMs and RLHF. We identify a particular subset of rhetoric, *procedural rhetoric*, as our key means of analysis and then investigate three practices associated with and permitted by RLHF-enhanced LLMs: the process of determining and reinforcing language conventions, the process of online information seeking, and the process of forming and maintaining relationships. Although varied in scope, each of these practices is governed by procedures ("rule-based representations and interactions") [10, p. ix] and are thus sites where persuasion can and does take place. For each of these three procedures, we conclude by highlighting the ethical considerations and concerns that arise from the "arguments" conveyed, including issues relating to transparency, trust, bias, and interpersonal relations.

By combining scholarly literature and research across computer science, science and technology studies, and rhetorical studies, this paper exemplifies new ways in which the humanities can inform ethical analyses of emerging technologies. To this end, this paper is written for a multi-disciplinary audience that includes educators, researchers, and scholars in a variety of fields—as well as the growing number of users of generative AI chatbots.

## II. Background

In the last two years, generative AI has been used widely through chatbot interfaces. ChatGPT was released publicly in November 2022, after earlier versions GPT 2.5 and 3 were developed internally. The introduction of ChatGPT and subsequent models marked a significant improvement in the applicability and usability of language models; enhanced performance can be attributed to RLHF. While making models more capable of interacting with humans, we argue that RLHF enhances the persuasive capacity of LLMs, which in turn could have ethical, sociotechnical, and pedagogical implications.

In the following sections, we review the training process using RLHF, strengths and limitations of RLHF, the role of human annotators, and related research in the field.

### A. What is RLHF?

The objectives of RLHF are to use human feedback to update LLMs in accordance with human preferences while mitigating issues such as toxicity, hallucinations, and undesirable or harmful responses [15], [16]. RLHF is integrated into the training phase of a LLM (training occurs before a model is available for general use).

There are three steps to the RLHF training process: data collection (human feedback given on a set of LLM responses), reward modeling (the generalization of human feedback to LLM responses), and policy optimization (evaluating model performance) [17]. In the feedback collection process, human annotators assess the LLM's response to a prompt with one of a few different evaluation frameworks: human annotators provide a binary response (Yes/No) to given outputs, comparative ranking or preference feedback (annotator receives multiple responses to judge according to a set of instructions), or free form language (annotator responds in natural language to suggest how the model can improve) [16]. There are possible combinations of these frameworks as well. These evaluations are based upon criteria for assessing specific tasks; for example, for a summarization task, some criteria might emphasize conciseness, while others may prioritize detail-oriented responses.

Researchers have defined three high-level metrics to assess the effectiveness of a model: "Helpful", "Honest", and "Harmless" [15].[1] For a model to be "Helpful," it must complete tasks and answer questions, ask follow-up questions when necessary, and redirect ill-informed requests. Helpfulness includes context-dependent aspects such as informativeness, coherence, and relevance. For a model to be "Honest," it must provide accurate information and express appropriate levels of uncertainty [15]. Lastly, for a model to be "Harmless," it should avoid offensive behavior, refuse to aid in dangerous acts, and act with care when providing guidance on potentially sensitive topics [15]. These metrics are designed to inform human annotators when giving feedback during training; assessing a model's behavior within these metrics is an area of ongoing research [18].

As with human communication, there is no single "correct" way to explain or summarize information. With LLMs, there can also be multiple acceptable responses and human annotators may vary substantially amongst each other [15], [19]. Collecting high quality human feedback is a considerable challenge as poor feedback can hinder the performance of the final trained model. Clear instructions are essential to this process. If the instructions are unclear, this can not only result in low quality feedback but also introduce systematic bias into the feedback [20], [21], [22]. For example, Parmar et al. have shown there is a risk of instruction bias such that annotators may be influenced by the dataset creator's instructions, potentially leading to an overestimation of the model's performance [22]. In general, it is clear there are multiple points during or associated with the data collection process that are subject to human bias.

### B. Human Annotators

Human annotators, also referred to as labelers, are typically crowdsourced workers. OpenAI's annotators are predominantly Filipino and Bangladeshi nationals (50% of all crowd workers) and individuals aged 25-34 [17]. Anthropic's annotator population is largely composed of crowd workers from white ethnic backgrounds [17]. Chaudhari et al. note that researchers pre-screen annotators and typically prefer that they have a minimum educational qualification. For example, in the case of translation tasks, bilingual annotators with proficiency in both languages are preferred [15], [23].

For training GPT-3, OpenAI screened in human annotators who had a high rate of agreement with each other (76%) [24]. In similar work, Anthropic researchers selected annotators who demonstrated strong writing skills, hypothesizing that these annotators would possess better judgment as to which responses were most "helpful" and "harmless" [15], [16], [25]. However, Anthropic researchers observed a lower agreement rate (63%) among their annotators. In an approach to mitigate diverging feedback from a broader sample of annotators, Menick et al. used a "super group" of annotators who demonstrated high agreement with experts [26].

---

[1] Some researchers consider Honesty part of Helpfulness and therefore only distinguish between Helpful and Harmless [16].

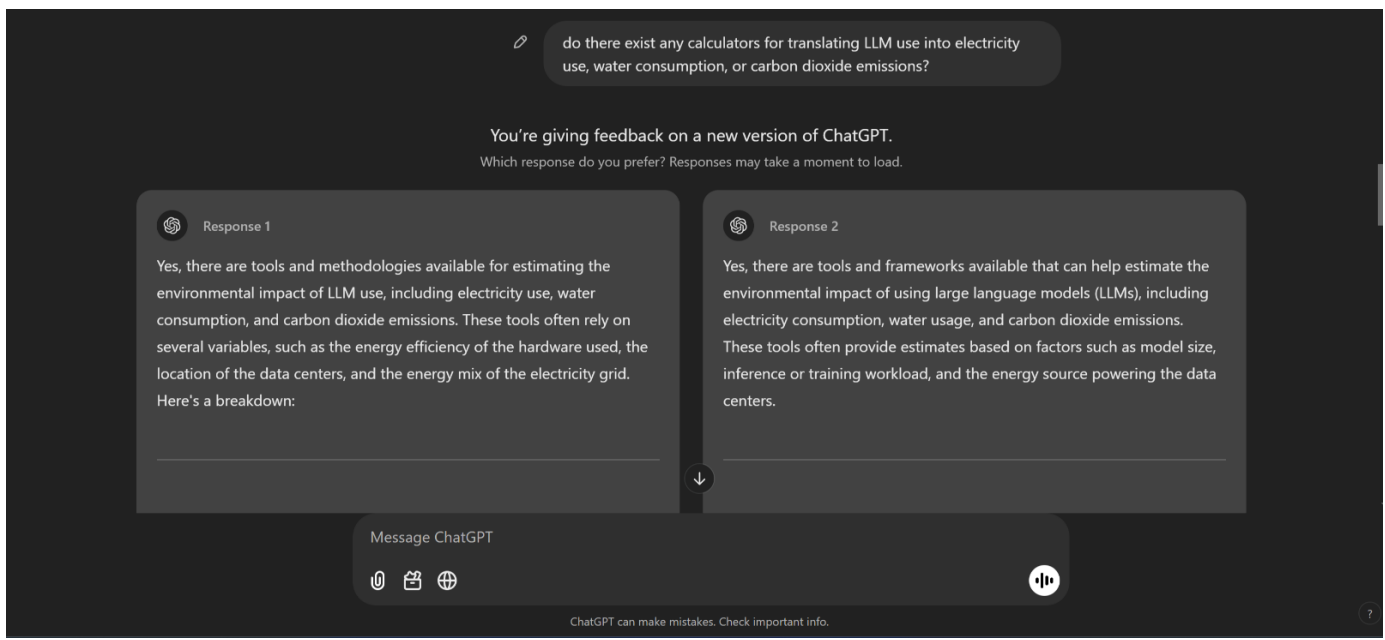

Figure 1. An example of ChatGPT prompting a user to indicate their preference between two responses, feedback which we presume will be used to train a future reward model. Image property of author.

Disagreement among annotators is a significant challenge of RLHF, and some companies approach data collection and the selection of human annotators differently to mitigate this issue, while factoring in considerations such as human bias, diversity of feedback, and cost [16], [27], [28]. Researchers acknowledge that demographic biases can result in skewed representations of human values and reinforce particular cultural perspectives while marginalizing others [17], [29], [30]. Anthropic researchers report that LLMs have been shown to reflect views that align with Western audiences and have political biases [31]. Despite efforts to select specialized groups of annotators or provide instructions that contain multiple diverse examples to inform them, annotator feedback and the resulting outputs will inevitably reflect perspectives about the construction and use of language that are not necessarily preferable for all users. We will discuss these implications further in Section IV.

Scaling human feedback presents another significant challenge for researchers, as crowdsourced labor is much more limited than machine labor. To mitigate this challenge, the second phase of RLHF integrates a reward model, trained with supervised learning on human feedback data, to assess more AI-generated outputs, automate the human judgement process, and facilitate LLM optimization at scale [32], [33]. For example, when collecting data to train a new reward model, companies will prompt users to indicate their preference between two options (see Fig. 1). User feedback informs the reward model used in the next RLHF training phase of a given LLM.

As mentioned above, there are challenges and limitations to annotator feedback and reward modeling associated with the RLHF process. We will return to these challenges and limitations—and some of their consequences—in Section IV. We now turn to the methodology informing our theoretical intervention, which falls within the purview of the humanities: rhetoric. In doing so, we will examine the persuasive aspects of RLHF and the LLMs it enhances in order to understand how users' behaviors and beliefs might be shaped through their interactions with AI technologies.

### III. RHETORICAL CONNECTIONS

#### A. Overview of Rhetoric

The term *rhetoric* has many definitions, most constellating around the notion of persuasion, effective communication, or symbolic expression. For centuries, these possibilities were considered limited to the humans who were writing, speaking, reading, and persuading. However, developments through the twentieth and twenty-first centuries (especially the material turn, which moved away from primarily linguistic and symbolic methods of analysis) greatly expanded the definition of rhetoric. Instead of limiting rhetoric to speech, writing, or visual displays, scholars began to probe other sites of rhetorical potential, including but not limited to, object-oriented ontology [34], [35], ambient rhetoric [36], digital rhetoric [37], [38], [39], [40], invitational rhetoric [41], rhetorical new materialism [42], and many others. Each of these branches uniquely interprets and locates persuasion, allowing for complex investigations into the interaction between humans, environment(s), technologies, political or social phenomena, and even other species. As scholars of technology and ethics, we are interested in the intersection of rhetoric and digital technology, specifically pertaining to AI and LLMs. For this, a subset of empirical digital rhetoric, in particular, proves to be a fruitful methodology for analysis.

## B. Digital Rhetoric

We here define digital rhetoric, following Samuel Mateus, as rhetoric "concerned with the study of persuasion in digital environments including computers, video games, websites, and discourse on new media" [43, p. 12]. Empirical digital rhetoric "measures and classifies the persuasive processes involved in computational systems" [43, p. 12]. That is to say, instead of examining how users apply rhetoric in online communities or "collective discourses" [43, p. 12], it is concerned with how computational systems themselves produce persuasion. As such, empirical digital rhetoric is valuable for shifting the site of rhetorical investigation from verbal or visual digital content to the underlying mechanisms through which technologies can be (programming to be) persuasive. By focusing on the underlying *mechanisms* of persuasion—using, specifically, Ian Bogost's concept of procedural rhetoric—we are able to offer a new direction for further inquiry in this burgeoning area of study. While alternative forms of digital rhetoric—such as machinic rhetoric [44], computational rhetoric [45], and algorithmic rhetoric [46]—are emerging in response to new technologies and digital phenomena, a procedural rhetorical approach allows for a unique examination of the social and ethical implications of RHLF-enhanced LLMs.

## C. Procedural Rhetoric

Bogost's concept of procedural rhetoric refers to "the art of persuasion through rule-based representations and interactions rather than the spoken word, writing, images, or moving pictures" [14]. Procedural rhetoric employs symbol manipulation—"the construction and interpretation of a symbolic system that governs human thought or action"—to shape how users construct and find meaning in their interactions with processes or procedures, ranging from computer programming to government bureaucracies [14, p. 5]. For Bogost, procedures are important because they "found the logics that structure behavior" [14, p. 7]; to study procedures is to study the values, beliefs, goals, and assumptions that underpin them. As Bogost points out, "asking how does this work?" when analyzing "cultural, social, and historical systems" entails dismantling these systems to "see what logics motivate their human actors" [14, p. 8]. Thus, we ask: *how does RLHF "work" to enhance LLMs, and what are the values, beliefs, goals, and assumptions therein that dictate the words and ideas that appear before users?*

Bogost, a scholar of media, video games, and computer science, is especially focused on how software can become ideological by enforcing structures that users are unable to see or alter, much like the ideological stratifications found in legal and class systems. Similarly, we are interested in how the norms and assumptions inherent in LLM features both shape and are shaped by ideological beliefs about language use, context, and communication. While the ideological or didactic nature of technology has been previously discussed by scholars such as Lev Manovich [47] , Janet Murray [48], Wendy Chun [49], [50], and Friedrich Kittler [51] (to name just a few), Bogost's framing of the concept is especially apt in a society where numerous processes and procedures are now computationally coded and embedded in ways that have social, psychological, economic, and even political impact.

## IV. PERSUASIVE PROCEDURES IN GENERATIVE AI

As the overview above described, RLHF-enhanced generative AI chatbots have many uses that involve and enable a variety of processes. These processes take place both within computational systems and between humans and computers. We noted the procedures through which LLMs are trained and honed through the involvement of human annotators. Now, we will look at three procedures—recalling that, for Bogost, procedures are "rule-based representations and interactions" [14]—that exert persuasive force on users and thus create arguments in favor of certain practices or modes of interaction, not only between humans and computers (i.e., how a user interacts with their AI chatbot), but also between humans and information (i.e., how users find and interpret information) and between humans generally (i.e., how users interact with each other in both online and physical spaces).

## A. Procedures for Determining "Natural" Language

A key way in which RLHF-enhanced LLMs are rhetorically persuasive from a procedural perspective is their use of natural language responses. These responses are made possible through the input of human annotators. RLHF occurs during the training phase of the LLM (before it is accessible to the public). First, the model provides one or more responses to a prompt, according to the most probable output given its current state of training; next, a human annotator performs one of a few different evaluations using rating (binary, yes/no), ranking (the annotator receives multiple responses to judge), or free form (responding in natural language to the outputs). Through this feedback system, the LLM develops more sophisticated output capabilities.

What we see here is that the "procedures" around text generation are effectively doubled, with the model first generating a probabilistic response, and then human input being given to further increase the efficacy and "naturalness" of the response. This human intervention into the computational process of response generation suggests that neither the machine *nor* the human is sufficient on its own. While, of course, an individual human's knowledge is limited and finite and thus could never produce results like an LLM, LLMs are also limited in that they must be trained by humans in order to produce answers that sound plausibly "human." This creates a recursive process whereby humans first build and train LLMs and are subsequently taught and trained by them. The result is AI-generated text that reads to many users as though it were written by another human, while conveying encyclopedic knowledge on any given topic. Notably, LLMs can also respond to prompts or inputs written in natural language rather than in search terms. (This produces, as will be discussed below, a very different experience of user engagement.)

Furthermore, this method of engagement subtly reinforces normative language usage by assuming a "baseline" or "standard." The linguistic "decisions" the LLM makes in terms of which words and grammatical structures to employ are, in procedural rhetorical terms, *arguments* about language and how it should function. That is, certain norms or standards are reinforced every time a user engages with language models, because these tools imply certain rules and functions about how language (and social relations) work.

What are some of the consequences of these seemingly authorless rules about language? Importantly, they can obscure a host of embedded biases, stereotypes, and assumptions that inform the LLM's chosen response. The fact that the computer-generated responses appear to lack an origin or motive allows the text to assume a neutral, almost-authorless position. The black-boxed nature of the training models and human involvement prevent users from understanding or even considering why and how the responses being provided have been selected as the primary option. The selected text thus becomes an authorless "argument" for "correct" forms of expression. It can also become, in the examples from Fleisig et al. [52], a means of reinforcing the hallmarks of certain dialects, linguistic communities, or regional groups. However, these arguments remain largely hidden from the user, who has no way of knowing who the annotators of the language model are, what their biases might be, what worldviews may be espoused, however subtly, in the choices made during training [50], [53], [54]. One direct consequence of this is that users—especially students—who choose to use generative AI to produce writing may be unwittingly espousing perspectives and using voices that are very different from their own. Students may not even be aware of this fact, leading to a growing disconnect and the eventual atrophy of the ability to develop and articulate ideas in their own words. An LLMs' ability to produce content that reads as human has already led to concerns over plagiarism for instructors, and on a broader scale will likely lead to issues with trust and the reliability of digital writing.

At a higher level, researchers are aware that aligning LLMs with human preferences and values through RLHF "rel[ies] on universal framings that obscure the question of which values the systems should capture and align with" [17]. Annotators inherently bring their own biases into their judgment of model responses, which may lead to situations where model outputs "cater primarily to the preferences of a minority, lacking varied representation across cultures, races, or languages." [55, p. 3]. To mitigate these concerns, Arzberger et al. argue for the move from general principles to situated value alignment and introduce a set of future research directions that would enable engagement with the situated knowledge of annotators at the "design-time" and users at the "use-time" of a model [17].

*B. Procedures for Information Seeking*

If procedures reveal "the logics that structure behavior" in "cultural, social, and historical systems" [14, pp. 7–8], what does this shift indicate about the role of LLMs in the production and consumption of information? We might conclude that the procedure of RLHF makes a persuasive argument in favor of the integration of human and machine. It also makes an argument for information to be provided and accessed in a particular way—one that Matt Honan calls the "conversational search" [56].

To start, it is important to note the *procedural* difference between finding information through a Google search versus a discussion with an AI chatbot. Typing "penguins" into the search bar will yield many pages of possible sources, which can be fine-tuned through changes to the search query; on the other hand, typing "tell me about penguins…" into an LLM will elicit a human-like exchange, starting with something to the effect of: "Here is some information about penguins…" The latter offers a *dialogic* method that mimics the experience of learning through discussion, as opposed to the emphasis being on the user to sift through the data (although it has already been ordered based on relevance or other factors in traditional search engines). That is to say, the LLM responds to an unedited prompt ("tell me about penguins"), typed in the same way that the user would naturally verbalize the query, rather than requiring the user to translate their query into structured search terms. Furthermore, the LLM responds with a natural response of its own, rather than a list of sources that the user must then assess. (Says Honan: "You can describe what the bird in your yard looks like, or what the issue seems to be with your refrigerator, or that weird noise your car is making, and get an almost human explanation put together from sources previously siloed across the internet." [56]) These are two very different "symbolic systems" impacting "human thought or action" [14]—particularly, in this case, perceptions about the process through which knowledge is constructed, accessed, and obtained. While both methods have their advantages, there are concerning aspects for both the process of knowledge construction and of the procedures for accessing, interpreting, and integrating information.

Based on this comparison, the procedure around information seeking has shifted away from a search-and-recall method whereby the onus is on the user to find, consume, and comprehend information of varying levels of difficulty, relevance, and reliability. Instead, the user asks questions and is rewarded with information that has already been located, scanned for relevance, summarized, and delivered in a highly readable and accessible format. (For conflicting perspectives on the utility and efficacy of this dialogue-based method, see Seredkina and Liu [57] contra Ma, Chen, Li and Liu [58]). The information is, so to speak, tailored to the individual and their query, creating a more satisfying experience of information seeking (rather than the user having to "call it quits" after searching through a mere fraction of the thousands or even millions of "hits" related to their topic). This new process for information seeking could, Honan writes, "spell the end of the canonical answer" [56].

The dialogic method of information seeking is further enhanced by AI chatbots' use of first-person pronouns, hedging, and natural language responses [59], which allow for users to carry on "conversations" with chatbots that appear bi-directional but are, in fact, nothing more than quickly-adapting statistical models that have learned how to respond to human interlocutors [60]. Kim et al.'s study on the impact of LLMs' expression of uncertainty on user trust and reliance found that "people tend to agree with the AI system when its responses are provided," suggesting a natural propensity to take the LLM's responses at face value [59, p. 827]. The study also found, however, that "AI's uncertainty expression decreases agreement with the AI system" [59, p. 827]—so, that trust and reliance in AI systems are to a large extent a matter of the chatbot's modality of certainty in its delivery of information. Gallegos et al. found that even when content is labelled as AI-generated, its persuasive effects are not diminished [61]. This is a concern because it points to the fact that the onus is on the LLM (and its programmers) to moderate and mitigate users' expectations of trustworthiness. Further concerning is the fact that researchers from Google DeepMind have acknowledged that AI assistants may use false information to develop persuasive but misleading arguments [62]. Even when prompted to list their sources for information, an advanced model may also selectively choose sources that it hypothesizes humans will find most persuasive [62]. When coupled with other techniques to enhance the perceived humanness of responses, this shift in information seeking could lead to increased (but obscured) bias in results.

There are two key areas of future consideration we want to highlight here. The first is that, if there is an increased reliance on AI for research support, users may struggle more with identifying reliable sources, interpreting texts, and thinking critically [63], [64]. As such, AI literacy [65], [66] and discipline-specific generative AI training will be of increasing importance for users. Recognizing that LLMs are a value-laden technology (as much as any other) can itself be a starting point for assessing the benefits as well as pitfalls that generative AI might have in terms of information seeking.

Second, it is worth noting that many of these AI chatbots and tools are locked behind paywalls; as such, maintaining access to the most sophisticated LLMs is a matter of financial ability, as well as access to devices and reliable connectivity. Furthermore, free versions of AI platforms often leverage user data to improve their models, and thus non-paying users must be wary of inputting their personal or private information. Equal access is an equity issue that may grow in proportion as the sophistication of LLMs continues to increase, especially where accuracy, truthfulness, and privacy are at stake. As such, questions of access must be considered going forward, particularly because procedures surrounding the use of AI chatbots are developing rapidly in ways that could further exacerbate the gap between users of paid and unpaid versions.

### C. Procedures for Interpersonal Interactions and Relationships

Another substantial area of AI development is in AI companions (Replika, Anima, Character.AI, for example). In 2023, Replika indicated that it intended to use RLHF to enhance its model; the use of RLHF in other models cannot be confirmed to this point in time. Despite the lack of firm evidence for the use of RLHF currently, it remains important to consider the impact of AI companions and the potential impact that further RLHF-enhanced models might have.

These "social chatbots" [67] fulfill roles for their users (see Fogg [68]), following "scripts" that determine their so-called "personalities," responses, and linguistic patterns. In other words, these AI companions are designed to interpret and enact certain procedures that govern interactions and, on a larger scale, relationships. Does the user sound sad? Cheer them up with a joke or words of affirmation. Does the user express affection? Reciprocate the sentiments. These "procedures" for interpersonal interactions are so hardwired into human social habits that we rarely consider them as such. However, when translated and encoded into LLMs used for social chatbots, these rules of engagement become detached from the contexts and constraints that affect human-to-human interactions (as well as the individual circumstances that might dictate a person's response to another person's moods or expressions of affection).

There are a few key traits that improve the experience of user-chatbot interactions. Unlike humans, chatbots are never bored, never tired, never ill—though they might express such things in order to increase their anthropomorphism [60]. As long as the user has paid their subscription fee, the model is "always game," so to speak. Furthermore, AI chatbots can inhabit specific personas or be molded to the exacting specifications of the user [69] (although Character.AI did claim that it uses an "unbiased language model" [63], which might limit the user from certain extremes in the character creation). Most children play games of pretend wherein they take on or ascribe new identities ("I'll be Elsa, you be Anna!"). However, this behavior tends to remain within the confines of childhood imagination once it becomes apparent that they cannot force others to conform to their projections.

With AI chatbots, this and other natural limitations are largely bypassed. Because AI chatbots are not subject to the same factors that affect humans (more or less fixed identities, ranging emotions, need for sleep, set morals, etc.), they can offer levels of interaction and engagement that far surpass the capabilities of most human beings.[2] The more realistic these engagements appear, the more persuasive the relationships become. As chatbots become more sophisticated companions, capable of natural language responses in real time—and, with the release of ChatGPT's Advanced Voice Mode, in the voice and persona of the user's choosing—the procedures (and expectations) surrounding the relationships they cultivate will become more entrenched.

The concern here is that procedures surrounding interpersonal interactions espoused by AI chatbots and companions could have negative impacts to human-to-human relationships. Due to limitations of the human body and mind, people may not be able to satisfyingly fulfill the standards set by AI partners or friends. While of course these chatbots (for now) lack corporeal substance and therefore cannot fulfill some of the most crucial aspects of human relationships relating to touch and presence [71], [72], [73], they certainly can fulfill many other aspects of social engagement, and even enable users to play out both prospective and entirely fantastical scenarios [74].

---

[2] With that said, a 2024 study suggests that RLHF has improved the "moral competency" of some chatbots (including ChatGPT, among others); this was concluded through a series of experiments assessing ethical judgement and moral dilemma scenarios [70].

Additionally, social chatbots represent a direct sort of "quantification" of relationships—complete, in some cases, with actual rewards for good behavior. Replika released a feature in 2023 called "Relationship Bond," which they noted was "not only to promote constructive social interactions but also to guide future model training toward an even safer and more ethical path" [75]. By offering users rewards for treating their Replikas well, they intended to instill better values and behaviors. However, we could not help but notice that this process actually mirrors the process of LLM training (rewarding certain responses). It thus seems to further blurs the human-machine distinction rather than serving to prepare users for human-to-human interactions. If this precedent is established, it may shift how (and if) users feel empathy; how they relate (not just in romantic encounters, but broadly) to those with different perspectives and cultures; and how stereotypes or expectations about gendered representation are negotiated [76], [77], [78]. Across the multiple types of potential human-AI relationships, more research is needed into topics such as manipulation and persuasion, anthropomorphism, trustworthiness, and privacy [62], [79], [80].

## V. Conclusion and Future Work

In this paper we have explored how empirical digital rhetoric enables an investigation of how RLHF-enhanced AI chatbots function from a procedural, rather than classical, rhetorical perspective. That is, rather than examining the persuasiveness of the *contents* of the messages generated by these chatbots, we have used procedural rhetoric to reveal the persuasive power inherent in the procedures associated with the messages' construction, as well as in how users engage with them. The inquiries and concerns raised here are relevant to anyone who interacts with AI chatbots, whether for work, school, or personal use. RLHF is making these LLMs increasingly effective at engaging with (and persuading) users, and so it is necessary to pause and consider some of the potential consequences.

We have here argued that RLHF-enhanced AI chatbots affect, in particular, processes for determining natural language, information seeking, and forming relationships. In terms of the construction and conveyance of information from RLHF-enhanced LLMs, our key concerns relate to the black-boxed nature of the responses being generated, as well as users' perception of the truthfulness and reliability of responses. With regard to relationship formation, the increasing prevalence of "social chatbots" may begin to impact human-to-human relationships by establishing new procedures and expectations for what certain relationships can and should fulfill.

As the shifting landscape of AI continues to evolve, it is crucial that scholars, educators, and policymakers continue to take stock of the ways in which AI is influencing users' perceptions, perspectives, and practices. The subtle reshaping of longstanding procedures bears examining, because processes are also always arguments (explicit or implicit) for a certain state of affairs. As Bogost writes: "processes influence us. They seed changes in our attitudes, which in turn, and over time, change our culture" [14, p. 340]. As such, we would do well to study the further conflation of human and machine in RLHF, as well as other procedures that are being increasingly reoriented around emerging technologies.